\documentstyle[11pt,newpasp,twoside,epsf]{article}
\def\edcomment#1{\iffalse\marginpar{\raggedright\sl#1\/}\else\relax\fi}
\reversemarginpar

\begin{document}
\title{Probing the X-ray Variability of X-ray Binaries}
 \author{Wenfei Yu}
\affil{Astronomical Institute ``Anton Pannekoek'', University of Amsterdam, Kruislaan 403, 1098 SJ, Amsterdam, The Netherlands}
\affil{Institute of High Energy Physics, Beijing, China (previous address)}

\begin{abstract}
Kilohertz quasi-periodic oscillations (kHz QPOs) has been regarded as 
representing the Keplerian frequency at the inner disk edge in the 
neutron star X-ray binaries. The so-called ``parallel tracks'' 
on the plot of the kHz QPO frequency vs. X-ray flux in neutron star 
X-ray binaries, on the other hand, show the correlation between the 
kHz QPO frequency and the X-ray flux on time scales from hours to 
days. This is suspected as caused by the variations of the mass 
accretion rate through the accretion disk surrounding the neutron star. 
We show here that by comparing the correlation between the kHz QPO 
frequency and the X-ray count rate on a certain QPO time scale observed 
approximately simultaneous in the Fourier power spectra of the X-ray 
light curve, we have found evidences that the X-ray flux of millihertz 
QPOs in neutron star X-ray binaries 
is generated inside the inner disk edge if adopting that the kilohertz 
QPO frequency is an orbital frequency at the inner disk edge. 

\end{abstract}

\section{Introduction}
Kilohertz quasi-periodic oscillations (kHz QPOs) in the 
X-ray flux of about 20 neutron star low mass X-ray binaries
(LMXBs) (see van der Klis 2000 for a recent review) are 
suggested to be the indicator of the orbital frequency 
at the inner disk edge (Strohmayer et al. 1996; 
Miller, Lamb \& Psaltis 1998; Titarchuk, Lapidus \& Muslimov 1998; 
Stella, Vietri \& Morsink 1999;). 
One of the evidence for this, in addition to the millisecond 
time scales themselves, is that there are correlations between 
the kHz QPO frequency and the X-ray flux in these neutron star 
X-ray binaries on time scales from hours to days (e.g. Yu et al. 1997; 
Mendez et al. 1998; Zhang et al. 1998). Especially in the 
two neutron star soft X-ray transients (SXTs) 
4U 1608-52 and Aquila X-1, the correlation holds in a large 
flux range up to an order of magnitude (Mendez 1999/astro-ph/9903469).   
The correlation forms the so-called ``parallel tracks'' in the 
plot of kHz QPO frequency vs. X-ray flux in ``atoll'' sources. 
For those ``Z'' sources the correlation in the plot of 
kHz QPO frequency vs. X-ray flux may appear as 
negative correlation, depending on the source branch it stays 
(Wijnands et al. 1997; Yu, van der Klis \& Jonker 2001). 
These seem consistent with the general idea that an increase 
of the mass accretion rate (thus an increase/decrease of the 
X-ray flux) will cause the inner disk edge at radius $R_{in}$ 
approach closer to the neutron star, which would introduce an 
increase of the orbital frequency $\nu_{O}$ at the inner disk edge, 
and vise versa. The response of the inner disk radius to the mass 
accretion rate through the disk has been widely accepted in both 
theory and observations. 

Under the assumption that the source movements on those correlation 
tracks are because of the variation of the mass accretion rate 
at the inner disk, and the kHz QPO frequency is an indicator of 
the orbital frequency at the inner disk edge (notice that the kHz QPO 
frequency is not necessary to be the orbital frequency itself), 
the correlation between the kHz QPO frequency and the X-ray flux on each 
correlation track then probably represents the correlation between 
the inner disk radius and the mass accretion rate through the inner disk. 
Thus the X-ray flux variabilities on each track which occurs on time scales 
of hours to days could be an accretion mode. Similar variability modes 
on shorter time scales of accretion origin will probably introduce the 
same correlation between the inner disk radius and the X-ray count rate 
on its time scale. This then leads to the same correlation between the 
kHz QPO frequency and the X-ray count rate as that on the long-term 
tracks, but on a shorter time scale. Usually the X-ray variability of 
the neutron star X-ray binaries contain a few simultaneous variability 
components, e.g. continuous noise components and QPOs, covering a 
large range of time scales from mHz to kHz. The correlation 
between the kHz QPO frequency and the X-ray count rate tracking 
a certain variability component then could be a diagnostic 
of the variability mode and answer whether it is the same 
as that forms the long-term correlation tracks. 

\section{Observations and Results}
The ``parallel tracks'' are usually observed in those ``atoll'' sources 
such as 4U 1608-52 (Yu et al. 1997; Mendez et al. 1998), Aquila X-1 
(Zhang et al. 1998; Mendez et al. 1999), and 4U 1728-34 (Mendez \& van der Klis  
1999). In those ``Z'' sources, the correlation between the kHz QPO 
frequency and the X-ray count rate is complicated. For example, on 
the Normal Branch (NB) of Sco X-1, the kHz QPO frequency appears 
anti-correlated with the X-ray count rate, but correlated to the 
inferred mass accretion rate (Yu, van der Klis \& Jonker 2001). 
  
\subsection{mHz QPO in the ``Atoll'' source 4U 1608-52}
4U 1608-52 is an X-ray burster and a soft X-ray transient (SXT). 
In the observations of its outburst decay by the {\it Rossi} 
X-ray timing explorer, kHz QPOs were observed showing its 
frequency correlated with the X-ray count rate on the 
time scales of hours to days but not longer (eg. Yu et al. 1997; 
Mendez et al. 1998). More recently, simultaneous with the 
kHz QPOs, millihertz QPOs at a frequency of 7-9$\times$${10}^{-3}$ Hz 
in the X-ray flux of 4U 1608-52 were observed. They are associated 
with the soft X-ray photons below 5 keV and probably have close 
relationship with the type I X-ray bursts. This leads to 
the speculation that the mHz QPO is caused by a mode of 
nuclear burning at the neutron star surface 
(Revnivtsev et al. 2001). The corresponding kHz QPO frequency 
evolution in 4U 1608-52 in response to the mHz pulses was 
studied with a peak-aligned method after removing the long-term 
trend by Yu \& van der Klis (2002). They found that the kHz QPO 
frequency is anti-correlated with the mHz QPO flux which can be 
tracked by the average mHz QPO count rate of 12 seconds. The kHz 
QPO frequency shift in response to the mHz QPOs is about 0.6 Hz. 
\begin{figure}
\plotone{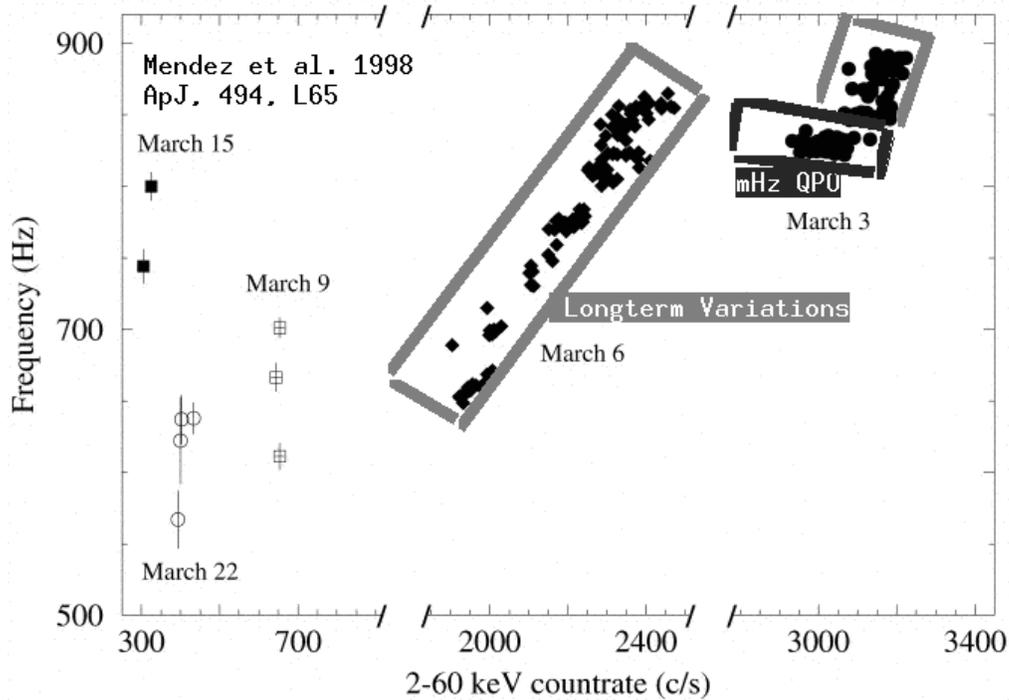}
\caption{The positive correlations between the kHz QPO 
frequency and the X-ray count rate on long time scales and on 
short time scales corresponding to mHz QPOs in 4U 1608-52 
(see also Yu \& van der Klis 2002). }
\end{figure}

\subsection{Variability modes in the plot of kHz QPO frequency 
vs. X-ray count rate}
To compare the X-ray count rate variation on a certain 
QPO time scales with that on longer time scales which 
forms ``parallel'' tracks in the plot of kHz QPO frequency 
vs. X-ray count rate, we show both the tracks and the 
branch corresponding to mHz QPOs showing the anti-correlation 
between the kHz QPO frequency and the X-ray count rate 
in Fig.1. On each correlation trackes, each data point 
represents the kHz QPO frequency and the X-ray count rate 
measured in 128 seconds. In the interval corresponding to 
the mHz QPOs, each data point represents that obtained in 
32 seconds. 

It is clearly visible that opposite to the overall positive 
correlation between kHz QPO frequency and X-ray count rate 
on longer time scales of hours, which forms two trackes 
from data on March 3, 1996 and March 6 , 1996, respectively, 
the kHz QPO frequency is anti-correlated with the X-ray 
count rate on the interval corresponding to the mHz QPOs.  
On this mHz QPO interval, the mHz QPO dominates the X-ray 
count rate variation. Thus from the plot of kHz QPO frequency 
vs. the X-ray count rate, we can discriminate the mHz 
QPO mode from that of the ``parallel'' tracks. 

\section{Discussion and Conclusion}
Here I show that the mHz QPOs in 4U 1608-52 can be identified 
in the plot of the kHz QPO frequency vs. the X-ray flux or 
by studying their correlation on a certain timescale corresponding 
to the mHz QPOs. This can be applied to the X-ray variability 
components we know in both neutron star X-ray binaries and 
possibly black hole X-ray binaries. The application to the noise 
components and other QPOs in neutron star X-ray binaries is 
straight forward and we may be able to determine the origin of 
those variability modes. 

W.Y would like to thank Prof.  Michiel van der Klis for comments 
which has been included in the above arguments. W.Y. also would 
like to acknowledge partial support by the Netherlands 
Organization for Scientific Research (NWO) under grant 614.051.002 
and partial support from NSFC (19903004). This work has made use of 
data obtained through the High Energy Astrophysics Science Archive 
Research Center Online Service, provided by the NASA/Goddard Space 
Flight Center.


\begin{references}
\reference Mendez, M. et al. 1998, \apj, 505, L23
\reference Mendez, M. 1999, astro-ph/9903469
\reference Mendez, M. \& van der Klis, M. 1999, \apj, 517, L51
\reference Miller, M. C., Lamb, F. K. \& Psaltis, D. 1998, \apj, 508, 791
\reference Revnivtsev, M. et al. 2001, \aa, 372, 138 
\reference Stella, L., Vietri, M.\& Morsink, S. M. 1999, \apj, 524, L63
\reference Strohmayer, T. et al., 1996, \apj, 469, L9
\reference Titarchuk, L., Lapidus, I. \& Muslimov, A. 1998, \apj, 499, 315 
\reference van der Klis, M. 2000, \araa, 38, 717
\reference Wijnands, R. et al. 1997, \apj, 490, L157
\reference Yu, W. et al. 1997, \apj, 490, L153 
\reference Yu, W., van der Klis, M. \& Jonker, P. 2001, \apj, 559, L29
\reference Yu, W. \& van der Klis, M. 2002, \apj, 567, L67
\reference Zhang, W. et al. 1998, \apj, 495, L9
\end{references}
\end{document}